\documentclass[%
twocolumn, 
superscriptaddress,
preprintnumbers,
 amsmath,amssymb,
 aps,
prb,
citeautoscript,
floatfix, 
]{revtex4} 

\usepackage{enumerate}
\usepackage{graphicx}
\usepackage{dcolumn}
\usepackage{bm}

\renewcommand{\imath}{i}

\showboxdepth=\maxdimen
\showboxbreadth=\maxdimen

\usepackage{xcolor}

\usepackage{xcolor}

\usepackage{xcolor}

\usepackage[normalem]{ulem}

\usepackage{xcolor}

\usepackage[normalem]{ulem}

\begin{document}

\preprint{Preprint version}

\date{\today}
\title{Zero-phase-difference Josephson current based on spontaneous symmetry-breaking via parametric excitation of a movable superconducting dot}

\author{A. M. Eriksson}
\author{A. Vikstr\"{o}m}
\affiliation{Department of Physics, Chalmers University of Technology, Kemig\aa{}rden 1, SE-412 96 G{\" o}teborg, Sweden}

\date{\today}

\begin{abstract}
Recent advances have attracted attention to non-standard Josephson junctions in which a supercurrent can flow despite zero phase difference between the constituent superconducting leads. 
Here, we propose a zero-phase-difference nanoelectromechanical junction which, in contrast to other considered systems, exhibits symmetry between leftward and rightward tunneling through the junction. 
We show that a supercurrent is nevertheless possible as a result of spontaneous symmetry breaking. 
In the suggested junction, the supercurrent is mediated by tunneling via a superconducting Cooper-pair box on a bendable cantilever. 
An alternating electric potential parametrically excites mechanical oscillations which are synchronized with charge oscillations of the box.
This leads to coherent transfer of Cooper pairs through the junction. 
The direction of the supercurrent is a result of spontaneous symmetry breaking and thus it can be reversed without changing the parameters. 
\end{abstract}



\maketitle

Josephson junctions exhibit well controllable quantum features and are therefore of interest to fundamental research~\cite{Friedman2000}. 
Josephson junctions have enabled state of the art sensor applications~\cite{Fagaly2006} and are promising as components in quantum information processing~\cite{Devoret2013}.
An ordinary Josephson junction consists of two superconductors separated by a thin potential barrier~\cite{Josephson1962}. 
If the superconductors are held at a non-zero superconducting-phase difference $\Delta \varphi$, tunneling through the barrier gives rise to a ground-state supercurrent. 
For the case of zero phase difference, $\Delta\varphi=0$, the tunneling has no preferred direction and the supercurrent is zero. 
However, a finite current can still exist if the symmetry between leftward and rightward tunneling is broken by other means. 
The junction then behaves as if it had an effective phase difference $\Delta\varphi+\varphi_0$. 
Many theoretical possibilities of such so-called ``$\varphi_0$-junctions'' have been proposed, such as multilayer ferromagnetic structures~\cite{Buzdin2008,Liu2010}, quantum point contacts~\cite{Zhang2016}, topological insulators~\cite{Dolcini2015}, quantum dots~\cite{Zazunov2009,Dell2007} and quantum wires~\cite{Yokoyama2014,Campagnano2015}. However, only very recently have one been realized experimentally by combining an external magnetic field and spin-orbit coupling~\cite{Szombati2016}.

    \begin{figure}\centering
    \includegraphics[width=\columnwidth]{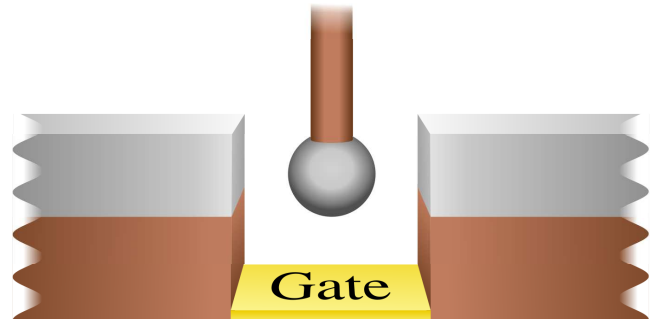}
    \caption{Schematic illustration of the system. A superconducting quantum dot (gray sphere) is attached to a bendable cantilever (brown rod) and positioned in the gap between two superconducting leads  (gray rectangular blocks). Cooper pairs can tunnel to and from the quantum dot from both sides of the gap. An alternating voltage applied to the gate (yellow) modulates the charging and decharging of the dot, leading to a parametric excitation of the cantilever bending mode.}\label{setUp}
    \end{figure} 

One suggestion for how to achieve a supercurrent between two superconductors with zero phase difference was put forth by Gorelik et. al.~\cite{Gorelik2001,Isacsson2002} They considered coherent transfer of Cooper pairs via a movable Cooper-pair box (CPB), a superconducting quantum dot~\cite{Bouchiat1998}. The CPB was modeled as a two-level system with a charge neutral state and a state with one excess Cooper pair. In their work, the CPB is artificially moved between two remote superconducting leads in a periodic manner. When the CPB is close to a lead, it can exchange Cooper pairs with it through tunneling. This puts the CPB  in a superposition of being charged and uncharged. While the CPB is moved towards the other lead, an electrostatic potential is applied. As a result, there is a change in the relative phase in the superposition of the charged and uncharged state. The applied potential thereby influences the future interaction of the CPB with the other lead. By reversing the electrostatic potential after each contact with a lead, the mirror symmetry is broken and an average Josephson supercurrent is established. In contrast to the supercurrent in a groundstate $\varphi_0-$junction, the supercurrent suggested by Gorelik et al. is a non-equilibrium phenomenon since it requires an explicitly time-dependent system.

In this letter, we propose a non-equilibrium nanoelectromechanical mechanism which coherently transfers Cooper-pairs between two superconductors with zero phase difference. In contrast to the work by Gorelik et al., the mirror symmetry in our system is not broken explicitly. Instead, we utilize spontaneous symmetry breaking via parametric excitation of the mechanical motion~\cite{Milton2013,Kang-Hun2006}. The supercurrent is established by the automatic synchronization of the mechanical oscillations and the effective charging and decharging of the CPB. 

In our nanomechanical junction, a CPB resides on a bendable cantilever which is inserted into the middle of the gap between two superconducting leads (fig.~\ref{setUp}). The bending mode of the cantilever allows the CPB to perform small oscillations between the leads. 
We will assume the leads to be bulk superconductors with zero phase difference, $\Delta\varphi=0$. 
Exchange of Cooper pairs between the superconducting leads is possible by tunneling via the CPB. The tunneling of Cooper pairs is assumed to not affect the superconducting bulk states. Further, the overlap integral between the CPB and a lead are assumed to decay exponentially with distance.  As a consequence, the tunneling coupling generates attractive 
forces between the CPB and the leads. The system is assumed to possess mirror symmetry (symmetry under the parity transformation $x\rightarrow-x$). In this case, the
forces toward the leads cancel each other in the middle of the gap, at the resting position of the cantilever. However, if the cantilever is bent so that the CPB moves slightly closer to either of the leads, the force toward that lead will dominate and soften the cantilever stiffness, i.e. decrease its spring constant. We will treat the electronic subsystem of the CPB as a two-level system with a charge neutral state, $|0\rangle$, and a state with one excess Cooper pair, $|1\rangle$.
The charging energy of the charged state is compensated by a static electrical field from a back-gate. Therefore the electronic ground state will be a superposition of the charged ($|1\rangle$) and uncharged ($|0\rangle$) state. 
Further, we apply a weak periodic electrical field with frequency $\Omega$ from the back-gate which modulates the energy of the charged state $|1\rangle$.

\begin{figure}\centering
\includegraphics[width=.7\columnwidth]{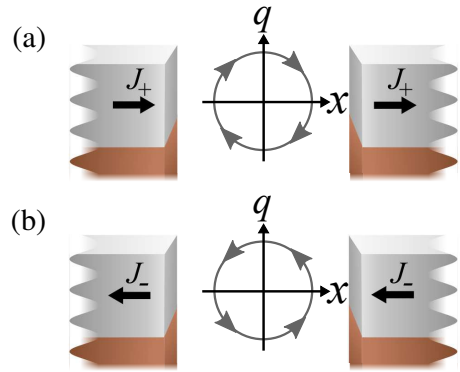}
\caption{The dynamics in charge ($q$) and position ($x$) space are automatically synchronized by the parametric excitation. The charge oscillations follow the driving field. The position has two possible oscillatory solutions which are out of phase with the driving field by approximately $\pm\pi/2$. Therefore the time-evolution in charge-position space is (a) clockwise or (b) counterclockwise. These ``chiral states'' coherently transfers Cooper pairs across the junction, leading to a supercurrent (a) to the right ($J_+$) or (b) to the left ($J_-$).\label{fig2}}
\end{figure}

Before giving a mathematical framework for the phenomenon, we will briefly give a physical picture of the dynamics. 
The periodic field applied to the electronic subsystem will be relayed through the tunneling coupling and modulate the cantilever stiffness. As we will show, the periodic modulation of the effective spring constant is able to parametrically excite the mechanical motion if the driving is strong enough to overcome the intrinsic mechanical damping~\cite{Landau1976}.
The strongest parametric excitation is achieved when the modulation frequency of the spring constant is close to two times the natural frequency $\omega_m$ of the mechanical oscillations.~\cite{Landau1976} 
Interestingly, in the suggested system, the strongest parametric excitation is achieved when the driving of the electronic subsystem is in resonance with the mechanical frequency $\Omega=\omega_m$. Hence, the cantilever oscillates with the same frequency $\Omega$ as the charging and decharging of the CPB. As a consequence, the driving field gives rise to synchronized oscillations of the mechanical position and the charge of the CPB (fig~\ref{fig2}).
The CPB will effectively be charging at one lead and decharging at the other lead, generating a supercurrent. 
The direction of the supercurrent is given by the relative phase between the charge and position oscillations. The charge oscillations simply follows the driving field. On the other hand, the parametric excitation leads to two oscillatory mechanical states with phase difference $\pi$. The time-evolution of the combined states is clockwise/counterclockwise in charge-position space (fig.~\ref{fig2}), and we will refer to these two electromechanical states as the ``chiral states'' of the system. 
The two chiral states are a result of spontaneous symmetry breaking and carry current in different directions. In an experimental situation, even weak interaction with the environment will occasionally induce transitions between the chiral states and thereby reverse the direction of the supercurrent. 
 
In the mathematical framework of the nanoelectromechanical system, we will model the cantilever mechanics as a quantum mechanical anharmonic oscillator with frequency $\omega_m$, effective mass $m$ and small Duffing nonlinearity $\eta$. The electronic subsystem of the CPB will be modeled as a charge qubit, as mentioned above. The Hamiltonian of the system takes the form  
\begin{align}
\label{ham1}
 \hat H=&\left(\frac{\hat p^2}{2m}+\frac{m\omega_m^2\hat x ^2}{2}+\frac{\eta}{4}\hat x^4\right)-2 eV_0 \cos(\Omega t)|1\rangle\langle1| \\ &-\frac{\hbar\omega_J}{2} \left(e^{-\hat x/\lambda} |1\rangle\langle0|+e^{\hat x/\lambda} |1\rangle\langle0|+\text{H.c.}\right). \nonumber
\end{align}
The first term describes the mechanical anharmonic oscillator with position (momentum) operator $\hat x$ ($\hat p$). The second term accounts for the alternating driving field with strength $eV_0\ll\hbar\omega_J$ which modulates the energy of the charged state. 
The second line originates from the tunneling of Cooper pairs where $\hbar\omega_J$ is the Josephson energy and $\lambda$ is the effective tunneling length. The two tunneling contributions describe Cooper-pair tunneling to the CPB from the left and right lead, respectively. 

An expression for the superconducting current can be obtained by considering the difference between rightward and leftward tunneling of Cooper pairs. 
The steady-state expectation value of the current will be an oscillatory function with period $T=2\pi/\Omega$. In order to investigate whether or not the system exhibits a direct current through the junction we average over one period. The time-averaged supercurrent is given by 
\begin{equation}
 \bar J=\frac{e\omega_J}{T}\int\displaylimits_{-T/2}^{T/2}\text{d}t\operatorname{Tr}\Bigg[i\Big(|0\rangle\langle1|-|1\rangle\langle0|\Big)\sinh\left(\frac{\hat x}{\lambda}\right)\hat\rho\Bigg]
\end{equation}
which we will refer to as the direct supercurrent.
 A more intuitive expression for the supercurrent is obtained by introducing the charge operator of the CPB $\hat q=-2e|1\rangle\langle1|$, using the Liouville von-Neumann equation $i\hbar \partial_t\hat\rho=[\hat H,\hat\rho]$ for the density operator $\hat\rho$, and integrating by parts. For small mechanical deflections, $\text{Tr}[\hat x^2\hat\rho]\ll\lambda^2$, we find (see Supplemental material~~\cite{supplemental}),
\begin{equation}\label{J}
 \bar J\approx - \frac{1}{T}\int_{-T/2}^{T/2}\text{d}t \operatorname{Tr}\left[\frac{\hat p}{2 m \lambda}\hat q\hat\rho\right].
\end{equation}
From this expression it is evident that a direct supercurrent can flow only if the charge and motion of the CPB are correlated in time. 

As we will show, the chiral states exhibit such correlation which results in a direct supercurrent through the junction although the phase difference is zero. To see this, it is convenient to use the Josephson representation $\langle0|=(1,-1)/\sqrt{2}$ and $\langle1|=(1, 1)/\sqrt{2}$ and write the Hamiltonian with Pauli matrices $\sigma_i$,
\begin{align}
\label{ham}
 \hat H=&\left(\frac{\hat p^2}{2m}+\frac{m\omega_m^2\hat x ^2}{2}+\frac{\eta}{4}\hat x^4\right) -\hbar\omega_J \hat{\sigma}_z\\&-\hbar \omega_J \epsilon \cos(\Omega t)\left(\hat I + \hat\sigma_x\right)-2\hbar\omega_J \sinh^2\left(\frac{\hat x}{2 \lambda}\right) \hat{\sigma}_z,\nonumber
\end{align}
where we have introduced the small driving-strength parameter $\epsilon=eV_0/\hbar\omega_J \ll 1$.
The large energy separation $\sim2\hbar\omega_\textrm{J}$ of the Josephson ground state $(|0\rangle+|1\rangle)/\sqrt{2}$ and excited state $(|0\rangle-|1\rangle)/\sqrt{2}$ is slightly modulated by the weak driving field and the electromechanical coupling described by the last term in eq.~\eqref{ham}. Note that there are no resonant transitions in the electronic subsystem since we assume $\Omega\sim\omega_m\ll\omega_J$. In order to calculate the supercurrent when the system is in one of the chiral states, we utilize the smallness of the driving $\epsilon$ and the mechanical deflection and calculate the effects of these perturbatively. As a first approximation, we disregard the electromechanical coupling. Hence, the density operator of the full system $\hat\rho$ is a product of the electronic and mechanical density operators: $\hat\rho\approx\hat{\rho}_e \otimes \hat{\rho}_m$. 
In this approximation, the supercurrent correlation separates to $ \operatorname{Tr}\left[\hat p\hat q\hat\rho\right]=\operatorname{Tr}\left[\hat p\hat\rho_m\right]\operatorname{Tr}\left[\hat q\hat\rho_e\right]$. The problem is then reduced to calculating the independent expectation values for the momentum and charge. We can perturbatively calculate the stationary oscillatory state $\hat{\rho}^\text{st}_e$ of the electronic subsystem under the influence of the driving field alone (see Supplemental Material~~\cite{supplemental}). 
 A small damping in the electronic system towards the unperturbed ground state removes the memory of the initial conditions and the average charge on the CPB is 
 \begin{equation}\label{q}
  \operatorname{Tr}[\hat{q} \hat{\rho}^\text{st}_e]\approx-e [ 1+2 \epsilon \cos(\Omega t) ] .
 \end{equation}
 
 \begin{figure}\centering
\includegraphics[width=\columnwidth]{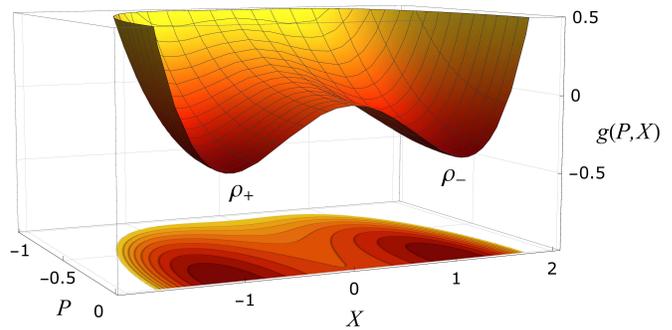}
\caption{The effective mechanical-energy landscape in the frame rotating with the electric driving field. 
Small damping will drag the system down to one of the two stable states $\rho_\pm$. 
In these states, the system performs synchronized electronic and mechanical oscillations which generate a supercurrent. 
The direction of the supercurrent is opposite for $\rho_+$ and $\rho_-$.
Stochastic fluctuations may therefore reverse the supercurrent by inducing transitions between the two states. 
The landscape is plotted for $\mu=0.2$. Note that $g( P, X)=g(-P,X)$. }  \label{effEnergy}
\end{figure}
 
 The oscillatory electronic state influences the mechanical dynamics via the tunneling coupling. The influence is taken into account by tracing the  Liouville von-Neumann equation $i\hbar\partial_t\hat\rho=[\hat H,\hat\rho]$ over the known state $\hat{\rho}_e^\text{st}$ of the electronic subsystem. We thereby obtain an effective equation for the mechanical density operator $\hat\rho_m=\text{Tr}_{e}[\hat\rho]$ under the influence of the electromechanical coupling. To lowest order in the driving parameter $\epsilon$, the effective mechanical equation becomes (see Supplemental Material~~\cite{supplemental})
\begin{align} \label{paramEx}
 i\hbar \partial_t {\hat\rho}_m&=\left[\hat{H}_\text{eff},\hat \rho_m\right],\\
 \hat{H}_\text{eff}&=\frac{\hat p^2}{2m}+\frac{m\tilde\omega_m^2\hat x ^2}{2}+ \frac{\eta}{4}\hat x^4+\epsilon^2 \frac{\hbar\omega_J}{8}\frac{\hat x^2}{\lambda^2}\cos(2\Omega t) ,\nonumber
\end{align}
where we have introduced the renormalized mechanical frequency $\tilde\omega_m^2=\omega_m^2 [1-\hbar \omega_J/(m \lambda^2)]$.
The effective mechanical eq.~\eqref{paramEx} describes the well known parametrically driven anharmonic quantum oscillator. The effective driving modulates the spring constant and pumps energy into the mechanical system. The pumping is most efficient at the resonance, $\Omega=\tilde\omega_m$. As the mechanical motion is pumped to higher amplitude, the effective separation of the mechanical energy levels will be modified by the anharmonic potential. As a consequence, the system is pushed out of resonance with the driving field and the mechanical amplitude will saturate. To calculate the stationary state of the mechanical system, it is convenient to transform to the rotating frame of the driving field. In the rotating wave approximation, the effective Hamiltonian is proportional to the time-independent energy landscape~~\cite{Dykman1998} 
\begin{equation}
\label{eq:potential}
 \hat{g}(\hat P,\hat X)=\frac{1}{4}(\hat P^2+\hat X^2)^2+\frac{1}{2}(1-\mu)\hat P^2-\frac{1}{2}(1+\mu)\hat X^2 .
\end{equation}
The new quantum variables $\hat P$ and $\hat X$ are defined in the appendix and the parameter $\mu=16 m\lambda^2\Omega\Delta/(\hbar\omega_J\epsilon^2)$, where $\Delta = (\Omega-\tilde{\omega}_m)/\tilde{\omega}_m$ is the detuning from resonance. 
When the system is close to the resonance, $-1<\mu<1$, the effective Hamiltonian has two minima (fig.~\ref{effEnergy}). Even very small damping of the mechanical motion will drag the system down to the minima corresponding to the states $\hat{\rho}_m^{\pm}$. These points correspond to the classical oscillatory states 
\begin{align}\label{p}
 \operatorname{Tr}[\hat p \hat{\rho}_m^{\pm}] &= m\Omega A \cos\left(\Omega t\pm\frac{\pi}{2}\right) , \\ 
 A&=\epsilon \sqrt{1+\mu} \sqrt{\frac{\hbar\omega_J}{6\lambda^2\eta}} \nonumber
\end{align}
with mechanical amplitude $A$ and phase difference $\pi$.

The direct supercurrent in the chiral states is obtained by combining the stationary electronic state $\hat{\rho}_e^\text{st}$ with the mechanical states $\rho^\pm_m$ calculated above. The two chiral states will carry supercurrent in opposite directions due to the flipped phase in the mechanical states. To lowest order in the small parameters we find the direct supercurrent \eqref{J} in the chiral states $\hat\rho_\pm=\hat{\rho}_e^\text{st}\otimes\hat{\rho}_m^{\pm}$ as
\begin{equation}
\bar J_\pm \approx - \int_{-T/2}^{T/2}\text{d}t \frac{\operatorname{Tr}[\hat p \hat{\rho}_m^{\mp}]
\operatorname{Tr}[\hat{q} \hat{\rho}^\text{st}_e]}{2 m \lambda T}
=\pm e\Omega\frac{A}{\lambda}\frac{eV_0}{2 \hbar\omega_J}\label{supercurrent}
\end{equation}
according to eqs.~\eqref{q} and \eqref{p}.
Hence, the direct supercurrent is proportional to the driving frequency $\Omega$, the driving field strength $e V_0$, and the amplitude of the mechanical oscillation $A$. 

A real system will inevitably be subject to damping toward and fluctuations around the chiral states. Although the noise levels are assumed to be low, the fluctuations may at rare instances lead to outbursts away from the double-well minima (fig.~\ref{effEnergy}). These outbursts may cause transitions between the chiral states and thereby reverse the direction of the supercurrent (fig.~\ref{fig2}). 
The fluctuations can have either a quantum or a classical origin. However, classical noise dominates in environments with temperature $k_B T_\text{env} \gg\hbar\omega_m$~\cite{Marthaler2006,Peano2012}. We consider such temperatures and assume that the most important noise source is the coupling of the mechanical degree of freedom to its environment.

To investigate the effect of mechanical damping and thermal fluctuations, we adopt the semi-classical model
\begin{equation}
\label{osc}
 \ddot x +\gamma \dot x +\omega_\textrm{m}^2x+\eta x^3=\xi(t)-\frac{1}{m}\frac{\partial}{\partial x}\text{Tr}\left[\hat H_\textrm{J}(x)\hat\rho_e\right],
\end{equation}
where $x$ is the classical position variable of the CPB, $\gamma$ is a small damping coefficient, $\xi(t)$ is a weak stochastic force and the last term describes an effective force from the electronic subsystem due to the semi-classical Josephson coupling $\hat H_J(x)=\hbar\omega_J\cosh(x/\lambda)\hat\sigma_z$. 
The semi-classical model reproduces the expression for the time-averaged supercurrent, eq.~\eqref{supercurrent}, as well as provides a criterion for mechanical excitation (see Supplemental Material~~\cite{supplemental}),
\begin{equation}\label{eq:excitation}
 \delta^2 > 4\left(\frac{\gamma^2}{\tilde{\omega}_m^2}+\Delta^2\right) ,\;
 \delta = \frac{1}{2} \left(\frac{e V_0}{\hbar \tilde{\omega}_J} \right)^2 \frac{\tilde{\omega}_J}{\tilde{\omega}_m} \left( \frac{a_0}{\lambda}\right)^2 ,
\end{equation}
where we have introduced the zero-point amplitude of vibrations $a_0=\sqrt{\hbar/(2 m\omega_m)}$ and neglected the weak stochastic force.

Next, we will estimate the average rate of transitions between the chiral states induced by the weak stochastic force $\xi(t)$. We assume that $\xi(t)$ is Gaussian noise with
\begin{equation}\label{eq:xicor}
 \langle \xi(t) \rangle = 0, \quad \langle \xi(t) \xi(t') \rangle = \frac{2 \gamma k_B T_\text{env}}{m} \delta(t-t') .
\end{equation}
We will consider the case when the excitation criterion \eqref{eq:excitation} is met and the driving is in resonance with the mechanical system, $\Delta=0$. 
Following Dykman et al.~\cite{Dykman1998}, the transition rate $\nu$ between the chiral states is given by $\nu=\omega_t \exp[-E_A/(k_B T_\text{env})]$, where $\omega_t$ is the attempt rate of transitions and the exponential factor is the success probability of each attempt. The activation energy of the transition is $E_A \approx [(4/\pi)-1] \varepsilon^2 m \omega_m^2 \hbar  \omega_J /(2 \lambda^2 \eta)$. 
The attempt rate $\omega_t\sim \omega_m \epsilon^2(1-\tilde\omega_m^2/\omega_m^2)/8$ is estimated from the effective parabolic potential at the bottom of each valley in the double-well in fig.~\ref{effEnergy}. Since $\omega_t \ll \omega_m$, transitions between the chiral states are rare events. 
Thus the dynamics resembles a telegraph process where the supercurrent switches between the values corresponding to the chiral states (eq.~\eqref{supercurrent}, fig.~\ref{fig2}). 
The possibility for a supercurrent in either direction for the same system parameters, with no phase difference between the leads, is in contrast to other non-standard Josephson~junctions~\cite{Baselmans1999,Linder2008}. 

To conclude, we  have proposed a  nanoelectromechanical system in which Cooper pairs are coherently transferred through a Josephson junction despite zero phase difference between the constituent superconducting leads. 
The phenomenon utilizes spontaneous symmetry breaking via parametric excitation of a  movable superconducting quantum dot. The parametric excitation results in synchronized mechanical and electrical oscillations corresponding to two symmetry-breaking chiral states. 
The two chiral states carry supercurrent in opposite directions. 
Controlled switching between the states may be used to reverse the direction of the supercurrent without changing the system parameters. 

 \begin{acknowledgments}
 We thank Leonid Gorelik for valuable discussions and the Swedish Research Council (VR) for funding. 
 \end{acknowledgments}
 
 \appendix
 
 \section{Transformation to the rotating frame}
 
 We transform the effective dynamical equation for the mechanical subsystem, eq.~\eqref{paramEx}, to the rotating frame of the driving field defined by 
 \begin{align}
  \hat U^\dagger_R(t)\hat x\hat U_R(t)&=C[\hat P\cos(\Omega t)-\hat X\sin(\Omega t)]\\
  \hat U_R^\dagger(t)\hat p\hat U_R(t)&=C m\Omega[\hat P\sin(\Omega t)+\hat X\cos(\Omega t)]
 \end{align}
 where $[\hat X, \hat P]=i \hbar / (C^2 m \Omega)$, $ C=\epsilon \sqrt{\hbar\omega_J/(6\lambda^2\eta)} $, and  
 \begin{equation}
  \hat U_R(t)=\exp\left[-\left(\frac{\hat p^2}{2 m}+\frac{1}{2}m\Omega^2 \hat x^2\right)\frac{i t}{\hbar}\right] .
 \end{equation}

In the rotating wave approximation, where rapidly oscillating terms are neglected, the effective Hamiltonian becomes 
\begin{equation}
 \hat U_R^\dagger(t)\hat{H}_\text{eff}\hat U_R(t)-i \hbar \hat U_R^\dagger(t) \dot{\hat{U}}_R(t) \approx \frac{\hbar^2 \omega_{J}^2 \varepsilon^4}{96 \eta \lambda^4} \hat{g}(\hat P,\hat X)
\end{equation}
where $\hat{g}(\hat P,\hat X)$ an effective time-independent potential~\cite{Marthaler2006} in the new quantum variables $\hat P$ and $\hat X$, given by eq.~\eqref{eq:potential}.


\end{document}